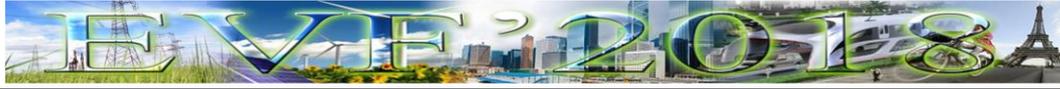

# A Single Phase DC/DC Microinverter with High efficiency and Harmonics Reduction using Passive Filters

**Lahcen El Iysaouy[1,2], Mhammed Lahbabi[2], Abdelmajid Oumnad[1]**

[1]Equipe de Recherche en Smart Communications - ERSC (ancien LEC)
EMI, UM5 ;Avenue Ibn Sina BP765 Agdal, Rabat, Morocco

lahcen.eliysaouy@um5s.net.ma ; aoumnad@emi.ac.ma

[2]LSSC, USMBA; B.P. 2202, Route d'Immouzzer, Fés, Maroc

lahbabi_m@yahoo.fr

**Abstract**

The photovoltaic microinverter based two-switch DC-DC flyback converters has been investigated [1,2]. The microinverter is characterized by the high performance and simple design. The effect of CL and LCL low pass filter on Total Harmonics Distortion (THD) of microinverter have been investigated. The results were found for the cases when CL and LCL output low pass filters were applied. The obtained results show that the efficiency and THD of the investigated photovoltaic microinverter based on the LCL output low pass filter is lower as compared to the case when the LC output low pass filter is utilized. The THD of microinverter with both output low pass filters is less than 5% respect the requirement of IEEE 1574-2014. Further, the LCL output low pass filter is used the THD decreases by 1% compared to the CL low pass filter.

## 1 Introduction

Development in PV microinverter increased due to their main role in low power PV purposes and to overcome the weaknesses created by the effects of shading in order to enhance their performance [3]. tow-stages is very forward topology and has a sampling manner of power conversion. In addition, the high switching frequency has to be considered in order to decrease the power losses and to ensure the reliable power quality of current injected in a grid. Besides, a single stage contains a DC-DC inverter, which boosts the voltage and ensures the galvanic isolation. Therefore, this microinverter belongs to the class of microinverter current source inverters type (CSIs) characterized by soft switching for the power MOSFETs in order to enhance the performance of the whole System. Therefore, this sort of CSIs has to be adopted by the filters in order the waveforms of the output current respect the terms of IEEE Std. 1574-2014. Thus, to control the ripple of output waveforms and to overcome total harmonic distortion (THD) that must be controlled by the output filter that considered as a critical element of microinverters. Further, the researchers have studied exceedingly the microinverter from the uncom-plicated filters, like inductor (L) filter, to the more complex filters like LCL, damped LCL with series damping resistor and another filter [1,4]. Although, the output CL filter for CSIs widely still used. Furthermore, the LCL filter has the best performance and overcome this issue of output harmonics which figures out with a combination of two inductors and capacitor see [5]. The main aim of this investigation CL and LCL filters is to find out the filters operate rightly with the grid connected. However, this investigation of the CL and LCL confirmed that the effectiveness of the result obtained within the LCL filter in microinverter and the quality of the signal injected into the grid-connected are better than find out by the CL filter if the parameters of transformer are chosen properly . Additionally, this paper complements the previous studies of CL and LCL filters.

## 2  System modeling

Fig. 1 illustrates a microinverter powered by dc voltage Vdc and filtered by a CL and LCL filters that will challenge in this paper constituted the inductors Lc, Lg, and capacitors C. The microinverter in Fig. 2 illustrates the topology of single stage microinverter [6]. Besides, the the high order frequency harmonics of the ac voltage filtered in using the CL and the LCL filters. Hence, the fig. 2 depict the filter CL and LCL where Li is the inductor side converter, Lg is the side of the grid connected, C is the capacitor and Vg is the grid connected voltage. The simulation has been performed for the parameters resumed in Table 1 using software Matlab/Simulink.

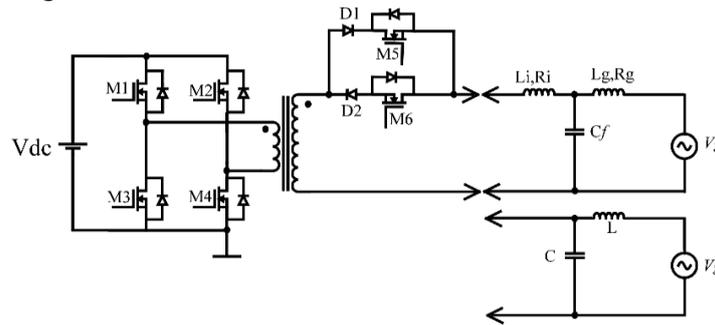

Figure 1: Single stage photovoltaic microinverter

Then the high order frequency harmonics of the formed AC voltage caused by the switching of microinverter switches are filtered by the LCL filter shows in fig. 1. However, the LCL filter is powered by the voltage Uc, where Lc is the inductor side converter, Lg is the side of the grid connected, C is the capacitor and Ug is the grid-connected voltage.

| Symbol | Definition | Value |
|--------|------------|-------|
| VDC | DC Voltage of photovoltaic module | 50 V |
| Vg | AC output voltage amplitude | 325 V |
| f | AC output voltage frequency | 50 Hz |
| fsw | Switching frequency | 25 kHz |
| Li | LCL filter inductor (inverter side) | 4.5 mH |
| Lg | LCL filter inductor (grid side) | 12 mH |
| L | CL filter inductor (grid side) | 5 mH |
| C | CL filter capacitor | 200 nF |
| Cf | LCL filter capacitor | 100 nF |

Table 1: Parameters of microinverter components

## 3  Results and discussion

The influence of flyback transformer primary winding resistance on the microinverter efficiency was investigated. The investigation was achieved for the cases when CL and LCL output low pass filter s are used. The results were performed at the 25 kHz switching frequency for 0.06 Ω, 0.12 Ω and 0.24 Ω flyback transformer primary winding resistances. The Fig. 3 present the microinverter efficiency dependences for the case of the microinverter with the CL output low pass filter and Fig. 4 with the LCL output low pass filter.

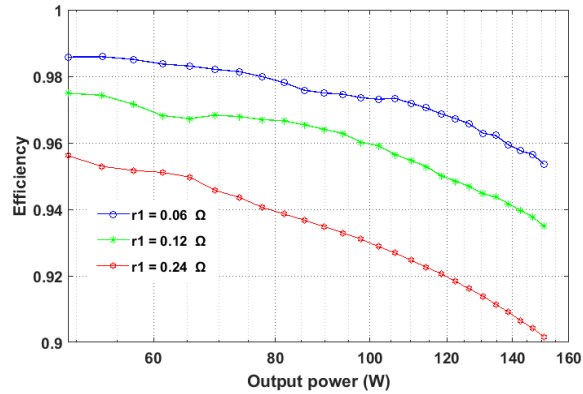

Figure 3: Flyback efficiency dependences on output power with the LCL low pass filter at 25 kHz switching frequency for various primary winding internal resistances r1 of flyback transformer.

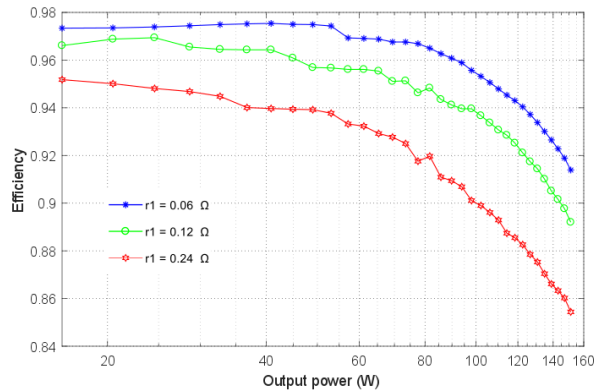

Figure 4: Flyback efficiency dependences on output power with the LCL low pass filter at 25 kHz switching frequency for various primary winding internal resistances r1 of flyback transformer.

It is seen that the decrease of primary winding resistance from 0.24 Ω to 0.06 Ω allows us to increase the efficiency at 100W output power from 0.93 to 0.97 for the microinverter with the CL output low pass filter and from 0.90 to 0.95 when the LCL output low pass filter is used. This fact shows that the reduction of flyback transformer primary winding resistance allows us significantly improve the efficiency of single stage photovoltaic microinverter.

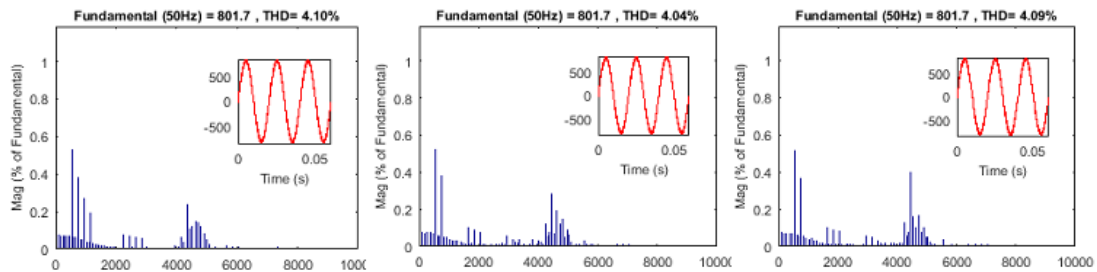

Figure 4: Flyback microinverter harmonics spectrum with the CL low pass filter at 25 kHz switching frequency for various primary winding internal resistances r1 of flyback transformer (a) Ron = 0.06 Ω. (b) Ron = 0.12 Ω. (c) Ron = 0.24 Ω.

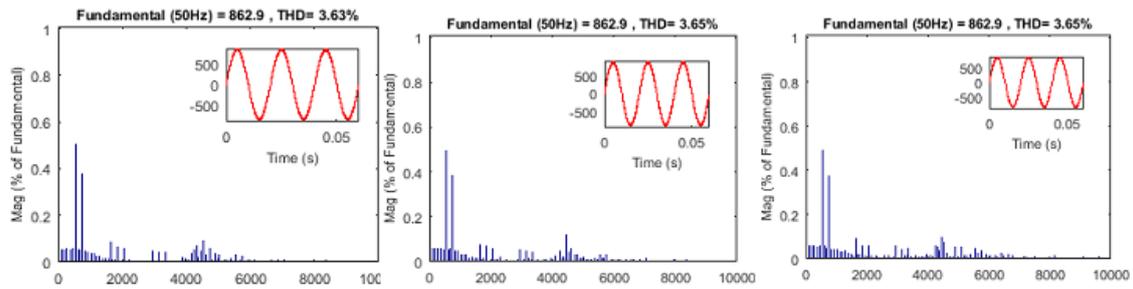

Figure 5: Flyback microinverter harmonics spectrum with the LCL low pass filter at 25 kHz switching frequency for various primary winding internal r esistances r1 of flyback transformer (a) r1 = 0.06 Ω. (b) r1 = 0.12 Ω. (c) r1 = 0.24 Ω.

The obtained result of THD harmonics are presented in Figs.4 and Fig. 5 for the microinverter with the CL and LCL output low pass filters, respectively. It is seen that the THD of flyback microinverter improved by 0.5% in case using the LCL low pass and reach a minumm value of THD 3.63% at r=0.06 Ω . Hence, we can figure out as a conclusion that flyback microinverter reach lower values of harmonics spectrum when LCL filter used respected the requirement of IEEE 1574-2014.

## 4   Conclusions

1. The efficiency of single stage photovoltaic microinverter based on the two-switch DC-DC flyback converters can reach 98% at 50 W and 96% at 150 W output power if the parameters of flyback transformer are chosen properly.
2. The reduction of flyback transformer primary winding resistance from 0.24 Ω to 0.06 Ω allows us to increase the efficiency at 100W output power from 0.93 to 0.97 of the microinverter with the CL output low pass filter and from 0.90 to 0.95 when the LCL output low pass filter is used.
3. The THD of two-switch DC-DC flyback converters can reach 3.63% at primary winding resistance r1= 0.06 if the parameters of low pass filter and flyback transformer and MOSFET transistors of switches are adapted well.
4. The utilization of the LCL low pass filter permits us to decrease the microinverter THD up to 1%.
6. The requirement of IEEE 1574 respected for both low pass filters.